\documentstyle[preprint,aps,epsf]{revtex}

\begin{document}

\title{Equation of state and magnetic susceptibility of
       spin polarized isospin asymmetric nuclear matter}

\author{Isaac Vida\~na and Ignazio Bombaci}

\address{Dipartimento di Fisica ``E. Fermi'', Universit\`a di Pisa and INFN Sezione di Pisa, 
Via Buonarroti 2, I-56127 Pisa, Italy}

\maketitle
     
\begin{abstract}

Properties of spin polarized isospin asymmetric nuclear matter are studied within the framework of the Brueckner--Hartree--Fock formalism.
The single-particle potentials of neutrons and protons with spin up and down are determined for several values of the neutron and proton spin polarizations  
and  the asymmetry parameter. It is found an almost linear and symmetric variation of the single-particle potentials as 
increasing these parameters. An analytic parametrization of the total energy per particle as a function of the asymmetry and spin polarizations is 
constructed. This parametrization is employed to compute the magnetic susceptibility of nuclear matter for several values of the asymmetry from neutron 
to symmetric matter. The results show no indication of a ferromagnetic transition at any density for any asymmetry of nuclear matter.  


\end{abstract}


\section{Introduction}
\label{sec:sec1}

Pulsars are believed to be rapidly rotating neutron stars endowed with strong 
magnetic fields \cite{pa67,go68}. Within the dipole magnetic model for pulsars 
the intensity of the surface (dipole) magnetic field $B_p$ can be inferred 
from the measured rotational period $P$ of the pulsar about its rotation axis 
and from its time derivative $\dot P = dP/dt$. For ``canonical'' pulsars 
{\footnote{Here we use the term ``canonical'' to distinguish these pulsars from 
the group of ``millisecond'' pulsars, for which $B_p \sim 10^{8}$--$10^{9}$~G, 
and from a possible new family of pulsars (the so-called ``magnetars'') 
having $B_p \sim 10^{15}$~G.}}
this method gives values for $B_p$ in the range $10^{12}$--$10^{13}$~G. 
Despite the great theoretical effort, there is no general consensus regarding the mechanism to generate such a strong magnetic field in a neutron 
star. The field could be a fossil remnant from the one of the progenitor star. In fact, assuming magnetic flux conservation during the birth of 
the neutron star, a magnetic field of $\sim 10^{12}$~G could originate from the collapse of a main sequence star with a typical surface magnetic 
field of 10--10$^2$~G. Alternatively, the field could be generated after the formation of the neutron star by some long-lived electric currents 
flowing in the highly conductive neutron star material.

There are strong theoretical and observational arguments which indicate a 
decay of the magnetic field during the ``life'' of a neutron star \cite{og69}. 
The physical processes which are responsible for the magnetic field evolution 
in neutron stars are not well understood, however, the observational data for 
pulsar distribution on the $P$-$\dot P$ plane can be reproduced by population 
synthesis codes (see e.g., ref. \cite{co01} and references therein quoted) 
in which the stellar magnetic field evolves according to the following 
exponential decay law \cite{bh91} 
\begin{equation}
   B(t) = B_{\infty} + (B_0 - B_{\infty}) e^{-t/\tau_B} \ ,
\label{eq:mag_field}
\end{equation}
where $\tau_B \sim 10^7$--$10^9$~yr is the field decay time and $B_{\infty} \sim 10^8$~G is a residue magnetic field.
If the residue field is permanent and is not generated by some dynamo mechanism, it could be produced by a spontaneous ferromagnetic
transition in the dense stellar core. Several authors have studied the possible existence of a ferromagnetic core in the liquid interior of 
neutron stars. First models, in which neutron star matter was approximated by pure neutron matter, were proposed just after the discovery of 
pulsars. Brownell and Callaway \cite{br69} and Rice \cite{ri69} considered a hard sphere gas model and showed that neutron matter becomes 
ferromagnetic at $k_F \approx 2.3$ fm$^{-1}$. Silverstein \cite{si69} and \O stgaard \cite{os70} found that the inclusion of long range attraction 
significantly increased the ferromagnetic transition density (e.g., \O stgaard predicted the transition to occur at $k_F \approx 4.1$ fm$^{-1}$ 
using a simple central potential with hard core only for the singlet spin states). Clark \cite{cl69} and Pearson and Saunier \cite{pe70} calculated 
the magnetic susceptibility for low densities ($k_F \leq 2$ fm$^{-1}$) using more realistic interactions. Pandharipande {\it et al.} \cite{pa72}, 
using the Reid soft-core potential, performed a variational calculation arriving to the conclusion that such a transition was not to be expected for 
$k_F \leq 5$ fm$^{-1}$. Early calculations of the magnetic susceptibility within the Brueckner theory were performed by B\"{a}ckmann and K\"{a}llman 
\cite{ba73} employing the Reid soft-core potential, and results from a correlated basis function calculation were obtained by Jackson {\it et al.} 
\cite{ja82} with the Reid $v_6$ interaction. A different point of view was followed by Vidaurre {\it et al.} \cite{vi84}, who employed 
neutron-neutron effective interactions of the Skyrme type, finding the ferromagnetic transition at $k_F \approx 1.73-1.97$ fm$^{-1}$. Marcos {\it 
et al.} \cite{ma91} have also studied the spin stability of dense neutron matter within the relativistic Dirac--Hartree--Fock approximation with 
an effective nucleon-meson Lagrangian, predicting the ferromagnetic transition at several times nuclear matter saturation density. On the contrary, no sign 
of such a transition has been found by Vida\~na {\it et al.} \cite{vi02}, who have recently studied the properties of spin polarized neutron matter 
within the Brueckner--Hartree--Fock approximation employing the realistic Nijmegen II and Reid93 interactions. In connection with the problem of neutrino
diffusion in dense matter, Fantoni {\it et al.} \cite{fa01} have recently employed the so-called auxiliary field diffusion Monte Carlo method (AFDMC)
using realistic interactions (based upon the Argonne $v_{18}$ two-body potential \cite{sm97} plus Urbana IX three-body potential \cite{pi98}), and have found
that the magnetic susceptibility of neutron matter shows a strong reduction of about a factor of $3$ with respect its Fermi gas value. They pointed out
that such a reduction may have strong effects on the mean free path of a neutrino in dense matter and, therefore, it should be taken into account in the
studies of supernovae and proto-neutron stars.

In all these studies neutron star matter was approximated by a pure neutron 
gas. Nevertheless, neutron star matter is also composed of protons, electrons,
muons and other exotic constituents \cite{pr97,he00}. The importance of the presence of a small number of protons in neutron matter for the spin stability 
was pointed out by Kutschera and W\'ojcik \cite{ku89}. These authors found that the ferromagnetic state, corresponding to completely polarized protons and 
weakly polarized neutrons, was energetically preferred over the nonpolarized one. Calculations performed by Bernardos {\it et al.} \cite{be95} for strongly 
asymmetric nuclear matter within the relativistic Dirac--Hartree--Fock approximation confirmed also that the presence of an admixture of protons favors the
ferromagnetic instability of dense matter. 

In this work we study the bulk and single-particle properties of spin polarized isospin asymmetric nuclear matter. Our calculations are 
based on the Brueckner--Hartree--Fock (BHF) approximation of the well known Brueckner-Bethe-Goldstone (BBG) theory of nuclear matter. 
To describe the bare nucleon-nucleon interaction we make use of the nucleon-nucleon part of the recent realistic baryon-baryon interaction (model NSC97e) 
constructed by the Nijmegen group \cite{st99}. We study the dependence of the single-particle potentials and the total energy per particle on the 
asymmetry parameter and neutron and proton spin polarizations. Further, we calculate the magnetic susceptibility for several values of the asymmetry
parameter and in particular we explore the possibility of a ferromagnetic transition in the high density region relevant for neutron stars.

The paper is organized in the following way. The formalism of the BHF approximation is briefly reviewed at the beginning of Sec. \ref{sec:sec2}, whereas the 
calculation of the magnetic susceptibility is presented at the end of it. Section \ref{sec:sec3} is devoted to the presentation and discussion of the results 
obtained for the single-particle potentials, the total energy per particle and the magnetic susceptibility. Finally, a short summary and the main conclusions 
of this work are drawn in Sec. \ref{sec:sec4}.


\section{Formalism}
\label{sec:sec2}

Spin polarized isospin asymmetric nuclear matter is an infinite nuclear system composed of four different fermionic components:
neutrons with spin up and spin down having densities $\rho_{n\uparrow}$ and $\rho_{n\downarrow}$ respectively, and
protons with spin up and spin down with densities $\rho_{p\uparrow}$ and $\rho_{p\downarrow}$.
The total densities for neutrons ($\rho_n$), protons ($\rho_p$), and nucleons ($\rho$) are given by:
\begin{equation}
    \rho_n = \rho_{n\uparrow}+\rho_{n\downarrow} \ , ~~~~~~~~~~~  
    \rho_p = \rho_{p\uparrow}+\rho_{p\downarrow} \ , ~~~~~~~~~~~
    \rho = \rho_{n} + \rho_{p}.
\label{eq:dens1}
\end{equation}
The isospin asymmetry of the system can be expressed by the asymmetry parameter $\beta  = (\rho_n - \rho_p) / \rho$,
while the degree of spin polarization of the system is characterized by the neutron and proton spin polarizations
$S_n$ and $S_p$, defined as
\begin{equation}
S_n=\frac{\rho_{n\uparrow}-\rho_{n\downarrow}}{\rho_n} \ , ~~~~~~~~~~~
S_p=\frac{\rho_{p\uparrow}-\rho_{p\downarrow}}{\rho_p} \ ,
\label{eq:spin_pol}
\end{equation}  
Note that the value $S_n=S_p=0$ corresponds to nonpolarized (i.e., $\rho_{n\uparrow}=\rho_{n\downarrow}$ and
$\rho_{p\uparrow}=\rho_{p\downarrow}$) matter, whereas $S_n=\pm 1(S_p=\pm 1)$ means that neutrons (protons) are totally polarized,
i.e., all the neutron (proton) spins are aligned along the same direction.

The single component densities are related to the total density and to the isospin and spin asymmetry parameters $\beta$, $S_n$, and $S_p$ via
the equations:  
\begin{equation}
\rho_{n\uparrow} = \frac{1+S_n}{2} \rho_n =
                    \frac{1+S_n}{2} \frac{1+\beta}{2} \rho \ ,
\label{eq:rhonu}
\end{equation}

\begin{equation}
\rho_{n\downarrow} = \frac{1-S_n}{2} \rho_n =
                     \frac{1-S_n}{2} \frac{1+\beta}{2} \rho \ ,
\label{eq:rhond}
\end{equation}

\begin{equation}
\rho_{p\uparrow} = \frac{1+S_p}{2} \rho_p =
                    \frac{1+S_p}{2} \frac{1-\beta}{2} \rho \ ,
\label{eq:rhopu}
\end{equation}

\begin{equation}
\rho_{p\downarrow} = \frac{1-S_p}{2} \rho_n =
                     \frac{1-S_p}{2} \frac{1-\beta}{2} \rho \ .
\label{eq:rhopd}
\end{equation}  
Finally the Fermi momentum $k_F^{\tau\sigma}$  is related to the corresponding partial density by $k_F^{\tau\sigma}=(6\pi^2\rho_{\tau\sigma})^{1/3}$,
with $\tau = n, p$ and $\sigma = \uparrow, \downarrow$.

Our calculations are based on the BHF approximation of the BBG theory extended to the case in which 
it is assumed that nuclear matter is arbitrarily asymmetric both in the isospin and spin degrees of freedom
(i.e., $\rho_{n\uparrow} \neq \rho_{n\downarrow} \neq \rho_{p\uparrow} \neq \rho_{p\downarrow}$). Therefore, our 
many-body scheme starts by constructing all the nucleon-nucleon $G$ matrices which describe in an effective way the 
interaction between two nucleons ($nn,np,pn$ and $pp$) for each one of the spin combinations ($\uparrow\uparrow, 
\uparrow\downarrow,\downarrow\uparrow$ and $\downarrow\downarrow$). The $G$ matrices can be obtained by solving the well 
known Bethe--Goldstone equation
\footnotesize
\begin{equation}
\begin{array}{c}
 \langle \vec{k}_1\tau_1\sigma_1;\vec{k}_2\tau_2\sigma_2|G(\omega)| \vec{k}_3\tau_3\sigma_3;\vec{k}_4\tau_4\sigma_4 \rangle =
 \langle \vec{k}_1\tau_1\sigma_1;\vec{k}_2\tau_2\sigma_2|v| \vec{k}_3\tau_3\sigma_3;\vec{k}_4\tau_4\sigma_4 \rangle \\
 \displaystyle{ +\sum_{ij}
 \langle \vec{k}_1\tau_1\sigma_1;\vec{k}_2\tau_2\sigma_2|v| \vec{k}_i\tau_i\sigma_i;\vec{k}_j\tau_j\sigma_j \rangle 
 \frac{Q_{\tau_i\sigma_i,\tau_j\sigma_j}}{\omega-E_{\tau_i\sigma_i}-E_{\tau_j\sigma_j}+i\eta}
 \langle \vec{k}_i\tau_i\sigma_i;\vec{k}_j\tau_j\sigma_j|G(\omega)| \vec{k}_3\tau_3\sigma_3;\vec{k}_4\tau_4\sigma_4 \rangle 
 } \ ,
\end{array}
   \label{eq:gmat}
\end{equation}
\normalsize
where $\tau_m$ and $\sigma_m$ indicate respectively the isospin and spin projections 
($n$$\uparrow$$(\downarrow)$,$p$$\uparrow$$(\downarrow)$) of the two nucleons in the initial, intermediate and final 
states, $\vec{k}_m$ are their respective linear momenta, $v$ is the bare nucleon-nucleon interaction, 
$Q_{\tau_i\sigma_i,\tau_j\sigma_j}$ is the Pauli operator which allows only intermediate states 
compatible with the Pauli principle, and $\omega$ is the so-called starting energy. 
Note that the $G$ matrices are obtained from a coupled channel equation. 
In practice a partial wave decomposition of the Bethe--Goldstone equation is performed, the Pauli operator and the energy denominator is replaced by an 
angle-averaged one, and the $G$ matrices are solved using relative and center-of-mass coordinates. 

The single-particle energy of a nucleon with momentum $k$ and spin projection 
$\sigma=\uparrow$($\downarrow$) is given by
\begin{equation}
E_{\tau\sigma}=\frac{\hbar^2k^2}{2m_{\tau}}+U_{\tau\sigma}(k) \ ,
\label{eq:spe}
\end{equation}
where the single-particle potential $U_{\tau\sigma}(k)$ represents the mean field ``felt'' by the nucleon due to its 
interaction with the other nucleons  of the system. In the BHF approximation $U_{\tau\sigma}(k)$ is calculated through the ``on-energy shell'' $G$ matrix 
and is given by
\begin{equation}
\begin{array}{c}
\displaystyle{U_{\tau\sigma}(k)
=\sum_{\tau'=n,p}\sum_{\sigma'=\uparrow,\downarrow}U_{\tau\sigma\tau'\sigma'}(k)} \nonumber \\
\displaystyle{=\sum_{\tau'=n,p}\sum_{\sigma'=\uparrow,\downarrow}\sum_{k' \leq k_F^{\tau'\sigma'}}\mbox{Re}
\langle \vec{k}\tau\sigma;\vec{k'}\tau'\sigma' |G(\omega=E_{\tau\sigma}+
E_{\tau'\sigma'})| \vec{k}\tau\sigma;\vec{k}'\tau'\sigma'\rangle_{\cal A}}  \ ,
\end{array}
\label{eq:spp}
\end{equation}
where $U_{\tau\sigma\tau'\sigma'}(k)$ is the  contribution to $U_{\tau\sigma}(k)$ due to the Fermi sea of nucleons
$\tau'\sigma'$. Note that a sum over the Fermi seas of neutrons and protons with spin up and down is performed and that the matrix elements are properly
antisymmetrized. We note also here that the continuous prescription has been 
adopted for the single particle potential when solving the Bethe--Goldstone 
equation. As shown by the authors of refs. \cite{so98,ba00}, the contribution 
to the energy per particle from three body clusters are diminished in this 
prescription. 

Once a self-consistent solution of Eqs.\ (\ref{eq:gmat}) and (\ref{eq:spp}) 
is obtained, the total energy per particle is easily calculated: 
\begin{equation}
\frac{E}{A}=
\frac{1}{A}\sum_{\tau=n,p}\sum_{\sigma=\uparrow,\downarrow}
\sum_{k \leq k_F^{\tau\sigma}}\left(\frac{\hbar^2k^2}{2m_\tau} +  
\frac{1}{2}U_{\tau\sigma}(k)\right) \equiv  \frac{T}{A} + \frac{V}{A}\ .
\label{eq:epp}
\end{equation}
This quantity is a function of $\rho_{n\uparrow}$, $\rho_{n\downarrow}$, $\rho_{p\uparrow}$
and $\rho_{p\downarrow}$ or, equivalently, of $\rho$, $\beta$, and $S_n$ and $S_p$.

The magnetic susceptibility $\chi$ of a system characterizes the response of 
this system to a magnetic field and gives a measure of the energy required 
to produce a net spin alignment in the direction of the field. In the case of 
nuclear matter it is defined by the $2\times2$ matrix 
\begin{equation}
\frac{1}{\chi}=
\left(\begin{array}{cc}
1/\chi_{nn} &
1/\chi_{np} \\
1/\chi_{pn}  &
1/\chi_{pp}
        \end{array}
  \right) \ ,
\label{eq:suscept}
\end{equation}
where the matrix elements $1/\chi_{ij}$ are given by
\begin{equation}
\frac{1}{\chi_{ij}}=\frac{\partial{\cal H}_i}{\partial{\cal M}_j} \ \ \ (i,j=n,p) \ .
\label{eq:suscept2}
\end{equation}
Here ${\cal M}_j$ is the magnetization of the species $j$ per unit volume 
\begin{equation}
{\cal M}_j=\mu_j(\rho_{j\uparrow}-\rho_{j\downarrow})=\mu_j\rho_jS_j\ ,
\label{eq:magneti}
\end{equation}
with $\mu_j$ the magnetic moment of the species $j$, and ${\cal H}_i$ is the magnetic field induced by the 
magnetization of the species $i$, which can be obtained from
\begin{equation}   
{\cal H}_i=\rho\frac{\partial (E/A)}{\partial {\cal M}_i} \ .
\label{eq:field}
\end{equation}  

Using Eqs. (\ref{eq:magneti}) and (\ref{eq:field}) the matrix elements $1/\chi_{ij}$ can be written as
\begin{equation}
\frac{1}{\chi_{ij}}=\frac{\rho}{\mu_i\rho_i\mu_j\rho_j}\left(\frac{\partial^2(E/A)}{\partial S_i \partial S_j}\right) \ ,
\label{eq:suscept3}
\end{equation}
where the second derivatives can be taken at $S_i=S_j=0$ if the magnetic field is assumed 
to be small.

It is convenient to study the magnetic susceptibility of the system in terms of the ratio 
det($1/\chi$)/det($1/\chi_F$), where $\chi_F$ is the magnetic susceptibility of the two component free Fermi gas. Writing the total energy 
per particle as a sum of the kinetic, $T/A$, and the potential, $V/A$, energy contributions, the ratio between both determinants 
can be written as
\begin{equation}
\frac{\mbox{det}(1/\chi)}{\mbox{det}(1/\chi_F)}
\Huge
=\left(1+\frac{\frac{\partial^2(V/A)}{\partial S_n^2}}{\frac{\partial^2(T/A)}{\partial S_n^2}}\right)
 \left(1+\frac{\frac{\partial^2(V/A)}{\partial S_p^2}}{\frac{\partial^2(T/A)}{\partial S_p^2}}\right)
-\left(\frac{\frac{\partial^2(V/A)}{\partial S_n \partial S_p}}{\frac{\partial^2(T/A)}{\partial S_n^2}}\right)
       \left(\frac{\frac{\partial^2(V/A)}{\partial S_p \partial S_n}}{\frac{\partial^2(T/A)}{\partial S_p^2}}\right) \ .
\label{eq:det}
\end{equation}

The stability of matter against spin fluctuations is guaranteed if det($1/\chi$)/det($1/\chi_F$)$>0$, indicating a change of sign of 
the ratio the onset of a ferromagnetic phase in the system.


\section{Results}
\label{sec:sec3}

\subsection{Single-particle potentials}
\label{sec:sec3.1}

In Fig.\ \ref{fig:fig1} we show the neutron $U_{n \uparrow}$, $U_{n \downarrow}$ (upper panels), and proton $U_{p \uparrow}$,
$U_{p \downarrow}$ (lower panels) single-particle potentials evaluated at $\rho=0.17$ fm$^{-3}$ for several values of $S_n$, $S_p$, and $\beta$.
Results for symmetric matter ($\beta=0$) are plotted on the left and middle panels, whereas results for asymmetric matter with $\beta=0.5$ are reported on 
the right ones. Solid lines show as a reference the results for nonpolarized matter ($S_n=S_p=0$), while dotted lines refer to spin polarized matter.
On the left panels we show the effects on $U_{\tau\sigma}(k)$ of a partial polarization ($S_n=0.75$) of the neutron component in
proton-unpolarized ($S_p=0$) symmetric matter. As it can be seen, the single-particle potentials of both species
split off when a partial polarization of the neutron spin is assumed. The single-particle potential for neutrons with spin up (down) becomes less (more) 
attractive with respect to $U_n(k)$ for nonpolarized nuclear matter. For the proton single-particle potential, we find an opposite qualitative trend
being $U_{p \uparrow}$ ($U_{p \downarrow}$) more (less) attractive than the proton single-particle potential for nonpolarized matter.
The additional partial proton polarization (middle panels) produces on 
$U_{\tau \sigma}(k)$ a reverse global effect with respect to the one produced 
by polarizing neutrons. This is particularly evident in the case of 
$U_{p \sigma}(k)$ (compare the results in the lower left and lower central 
panels). Finally, to obtain the results presented in the right panels of 
Fig.\ \ref{fig:fig1}, we further introduce an isospin asymmetry ($\beta = 0.5$) 
in the system. Notice that the isospin asymmetry generates a splitting of the 
neutron and proton single particle potentials even in the case of 
spin-unpolarized matter \cite{bl91}.  

The splitting, $U_{\tau\uparrow}$--$U_{\tau\downarrow}$, in the neutron and proton single-particle potentials can be mainly ascribed to two  
reasons: (i) the change in the number of pairs which the nucleon under consideration $|k,\tau,\sigma\rangle$ can form with the remaining nucleons
$|k^\prime \leq k_F^{\tau^\prime \sigma^\prime},\tau^\prime,\sigma^\prime\rangle$ of the system as nuclear matter is spin polarized, and (ii) the 
spin dependence (and isospin dependence for asymmetric spin polarized matter) of the nucleon-nucleon $G$ matrices in the spin polarized nuclear medium (see 
Eq. (\ref{eq:gmat})). 

In the general case of spin polarized asymmetric matter, the effect of the 
neutron and proton spin polarizations and the isospin asymmetry on the single 
particle potentials can be clarified by considering their partial 
contributions $U_{\tau\sigma\tau^\prime\sigma^\prime}(k)$ in such a way as to 
single out explicitly the dependence on their respective phase space: 
\begin{equation}
U_{n\uparrow} =  U_{n \uparrow n \uparrow} + U_{n \uparrow n \downarrow}
               + U_{n \uparrow p \uparrow} + U_{n \uparrow p \downarrow}
 = g_{n \uparrow n \uparrow}   \rho_{n \uparrow}
 + g_{n \uparrow n \downarrow} \rho_{n \downarrow}
 + g_{n \uparrow p \uparrow}   \rho_{p \uparrow}
 + g_{n \uparrow p \downarrow} \rho_{p \downarrow} \ ,
\label{eq:nu}
\end{equation}
\begin{equation}
U_{n\downarrow} =  U_{n \downarrow n \uparrow} + U_{n \downarrow n \downarrow}
                 + U_{n \downarrow p \uparrow} + U_{n \downarrow p \downarrow}
 = g_{n \downarrow n \uparrow}   \rho_{n \uparrow}
 + g_{n \downarrow n \downarrow} \rho_{n \downarrow}
 + g_{n \downarrow p \uparrow}   \rho_{p \uparrow}
 + g_{n \downarrow p \downarrow} \rho_{p \downarrow} \ ,
\label{eq:nd}
\end{equation}
\begin{equation}
U_{p\uparrow} =  U_{p \uparrow n \uparrow} + U_{p \uparrow n \downarrow}
               + U_{p \uparrow p \uparrow} + U_{p \uparrow p \downarrow}
 = g_{p \uparrow n \uparrow}   \rho_{n \uparrow}
 + g_{p \uparrow n \downarrow} \rho_{n \downarrow}
 + g_{p \uparrow p \uparrow}   \rho_{p \uparrow}
 + g_{p \uparrow p \downarrow} \rho_{p \downarrow} \ ,
\label{eq:pu}
\end{equation}
\begin{equation}
U_{p\downarrow} =  U_{p \downarrow n \uparrow} + U_{p \downarrow n \downarrow}
                 + U_{p \downarrow p \uparrow} + U_{p \downarrow p \downarrow}
 = g_{p \downarrow n \uparrow}   \rho_{n \uparrow}
 + g_{p \downarrow n \downarrow} \rho_{n \downarrow}
 + g_{p \downarrow p \uparrow}   \rho_{p \uparrow}
 + g_{p \downarrow p \downarrow} \rho_{p \downarrow} \ ,
\label{eq:pd}
\end{equation}
where $g_{\tau\sigma\tau^\prime\sigma^\prime}$ is the average value of the matrix element
$\langle \vec{k}\tau\sigma;\vec{k'}\tau'\sigma'|G|\vec{k}\tau\sigma;\vec{k'}\tau'\sigma' \rangle_{\cal A}$
in the Fermi sphere with radius $k' \leq k_F^{\tau^\prime\sigma^\prime}$ and the density factor $\rho_{\tau^\prime\sigma^\prime}$ in each term
arises from the integral over the corresponding Fermi sea. 
\par
For small values of the asymmetry parameter ($|\beta| << 1$) and the spin 
polarizations ($|S_n|, |S_p| << 1$) one can neglect the dependence on 
$\beta, S_n$ and $S_p$ of the average $G$ matrices $g_{\tau\sigma \tau^\prime\sigma^\prime}$  
assuming  $g_{\tau\sigma \tau^\prime\sigma^\prime} \sim 
           g_{\tau\sigma \tau^\prime\sigma^\prime}(k,\rho)$ and    
\begin{equation}
\begin{array}{c}
g_{n\uparrow n\uparrow} \approx  g_{n\downarrow n\downarrow} \approx
g_{p\uparrow p\uparrow} \approx  g_{p\downarrow p\downarrow} \equiv g_1 \ ,
\nonumber \\ 
g_{n\uparrow n\downarrow} \approx g_{n\downarrow n\uparrow} \approx
g_{p\uparrow p\downarrow} \approx  g_{p\downarrow p\uparrow} \equiv g_2 \ ,
\nonumber \\
g_{n\uparrow p\uparrow} \approx g_{n\downarrow p\downarrow} \approx
g_{p\uparrow n\uparrow} \approx g_{p\downarrow n\downarrow} \equiv g_3 \ ,
\nonumber \\
g_{n\uparrow p\downarrow} \approx g_{n\downarrow p\uparrow} \approx
g_{p\uparrow n\downarrow} \approx g_{p\downarrow n\uparrow} \equiv g_4
\nonumber \ .
\end{array}
\label{eq:g1234}
\end{equation}  
Clearly, even under these assumptions, in the most general case $g_1 \neq g_2 \neq g_3 \neq g_4$ because they receive contributions from 
different spin and isospin channels. Whereas $g_1$ receives contributions only from the spin and isospin triplet ($S=1,T=1$) channels, 
$g_2$ has in addition contributions from the spin singlet ones, $g_3$ from channels with $S=1$ and $T=0,1$, and $g_4$ from channels with 
$S=0,1$ and $T=0,1$. 

Using Eq. (\ref{eq:g1234}) the single-particle potentials can then be 
rewritten in terms of the average $G$ matrices $g_{\tau\sigma\tau^\prime\sigma^\prime}$ as:
\begin{equation}
\begin{array}{r}
U_{n\uparrow} \approx  \displaystyle{\frac{\rho}{4}}
\left[ (g_1+g_2+g_3+g_4) + (g_1+g_2-g_3-g_4)\beta \right. \\
\left. +(g_1-g_2)(1+\beta)S_n + (g_3-g_4)(1-\beta)S_p \right] \ ,
\end{array}
\label{eq:nu2}
\end{equation}
\begin{equation}
\begin{array}{r}
U_{n\downarrow} \approx  \displaystyle{\frac{\rho}{4}}
\left[ (g_1+g_2+g_3+g_4) + (g_1+g_2-g_3-g_4)\beta \right. \\
\left. -(g_1-g_2)(1+\beta)S_n - (g_3-g_4)(1-\beta)S_p \right] \ ,
\end{array}
\label{eq:nd2}
\end{equation}
\begin{equation}
\begin{array}{r}
U_{p\uparrow} \approx  \displaystyle{\frac{\rho}{4}}
\left[ (g_1+g_2+g_3+g_4) - (g_1+g_2-g_3-g_4)\beta \right. \\
\left. +(g_3-g_4)(1+\beta)S_n + (g_1-g_2)(1-\beta)S_p \right] \ ,
\end{array}
\label{eq:pu2}
\end{equation}
\begin{equation}
\begin{array}{r}
U_{p\downarrow} \approx \displaystyle{\frac{\rho}{4}}
\left[ (g_1+g_2+g_3+g_4) - (g_1+g_2-g_3-g_4)\beta \right. \\
\left. -(g_3-g_4)(1+\beta)S_n - (g_1-g_2)(1-\beta)S_p \right] \ ,
\end{array}
\label{eq:pd2}
\end{equation}
where Eqs. (\ref{eq:rhonu}), (\ref{eq:rhond}) (\ref{eq:rhopu}) and (\ref{eq:rhopd}) have been used to write $\rho_{\tau\sigma}$ in terms of $\rho$, $\beta$, 
$S_n$ and $S_p$. These equations show explicitly the dependence of the single-particle potentials on the spin polarizations and isospin 
asymmetry. This dependence is tested in Fig.\ \ref{fig:fig2}, where the value at $k=0$ of the single-particle potentials $U_{n\uparrow}, U_{n\downarrow}, 
U_{p\uparrow}$ and $U_{p\downarrow}$ at $\rho=0.17$ fm$^{-3}$ is plotted as a function of neutron (left panels) and proton (right panels) spin polarizations 
for two values of the asymmetry parameter: $\beta=0$ (upper panels) and $\beta=0.5$ (lower panels). In order to make more clear the discussion $S_n (S_p)$ 
is taken equal to 0 on the right (left) panels. The above equations predict a linear and symmetric variation of the single-particle potentials on $\beta$, 
$S_n$ and $S_p$. This prediction is well confirmed from the microscopic 
results reported on Fig.\ \ref{fig:fig2}, although deviations from this 
behavior are found at higher values of the asymmetry and spin polarizations. 
These deviations have to be associated to the dependence on $\beta, S_n$ 
and $S_p$ of the average $G$ matrices $g_{\tau\sigma \tau^\prime\sigma^\prime}$   
which has been neglected in the present analysis (see Eq. (\ref{eq:g1234})). 
From the above expressions it can be seen that under charge exchange 
(i.e., doing the changes $\beta \rightarrow -\beta, S_n \rightarrow S_p, S_p \rightarrow S_n$) the role of neutrons and 
protons with the same spin projection is exactly interchanged. This can be clearly seen by comparing the left and right top panels of the figure, which 
correspond to the particular case of symmetric matter. It is also clear that neutrons (protons) with spin up and down interchange their roles when a global 
flip of the spins is performed (i.e., changing $S_n$ by $ -S_n$ and  $S_p$ by $-S_p$).

\subsection{Energy per particle}
\label{sec:sec3.2}

The total energy per particle of neutron, asymmetric, and symmetric matter is shown 
on the left, middle and right panels of Fig.\ \ref{fig:fig3}, respectively. In the three panels solid lines show results for 
nonpolarized matter, whereas those for totally polarized matter are reported by dotted lines. Note that in totally 
polarized asymmetric and symmetric matter, we have distinguished two possible orientations of the neutron and proton spins: that in 
which neutron and proton spins are aligned along the same direction (i.e., $S_n=S_p=\pm1$), and that in which neutron and proton spins 
are orientated along opposite directions (i.e., $S_n=\pm 1, S_p=\mp 1$). As can be seen from the figure, totally polarized matter is always 
more repulsive than nonpolarized matter in all the density range explored for any value of the asymmetry parameter. Note also that the case in which neutron 
and proton spins are parallelly orientated is less repulsive than the case 
in which they have an antiparallel orientation. 
To highlight the effects of the nuclear interaction on the variation of $E/A$  
as nuclear matter is polarized, we plot in Fig.\ \ref{fig:fig4} the kinetic 
(upper panels) and potential (lower panels) energy per particle for 
nonpolarized  and totally polarized matter. 
The kinetic energy contribution of totally polarized 
neutron matter, asymmetric or symmetric matter is always larger than the 
corresponding one of nonpolarized matter, simply due to the fermionic nature of nucleons.   
Also, the potential energy contribution is always more repulsive in the totally polarized case. This can be understood by considering separately the 
contribution to the potential energy per particle of the spin singlet and spin triplet channels. In Tables \ref{tab:tab1} and \ref{tab:tab2} we show these 
contributions for nonpolarized and totally polarized neutron and symmetric matter at densities $\rho=0.17$ fm$^{-3}$ and $\rho=0.4$ fm$^{-3}$, respectively. 
Note, firstly, that an important amount of binding is lost in totally polarized neutron matter and in totally polarized symmetric matter with neutron 
and proton spins parallelly orientated due to the absence in these cases of the contribution of the spin singlet channels. Note also, in the totally 
polarized symmetric case, that when the orientation of neutron and proton spins is antiparallel this contribution, although it is present, is much less 
attractive or even repulsive (see Table \ref{tab:tab2}) than the 
corresponding one of nonpolarized symmetric matter. 
In addition, in all cases the contribution from spin triplet channels in 
totally polarized matter is always less attractive (or even repulsive at 
high density) than the corresponding one to nonpolarized matter.  
In particular, it is mainly this contribution the one which explains the 
difference between the energies of totally polarized symmetric matter with 
a parallel and an antiparallel orientation of neutron and proton spins, 
being in the antiparallel case less attractive or more repulsive. 
An interesting conclusion which can be inferred from our microscopic 
calculations is that a spontaneous transition to a spin polarized state, 
i.e., to a so-called ferromagnetic state, of nuclear matter is not 
to be expected for all the possible isospin asymmetries ranging from symmetric 
to pure neutron matter. 
If such a transition would exist, a crossing of the energies of the totally polarized and nonpolarized 
cases would be observed in neutron, asymmetric or symmetric matter at some density, indicating that the ground state of the system would be ferromagnetic 
from that density on. As can be seen in Fig.\ \ref{fig:fig3}, there is no 
sign of such a crossing and, on the contrary, a spin polarized state becomes 
less favorable as the density increases.  

It would be interesting to characterize in a simple analytic form the 
dependence of the total energy per particle on the asymmetry and spin 
polarizations. The kinetic energy contribution is already analytic and well 
known. It is given by
\begin{equation}
\begin{array}{r}
\displaystyle{\frac{T}{A}(\rho,\beta,S_n,S_p)=\frac{3}{5}\frac{\hbar^2k_F^2}{2m}\frac{1}{4}}
\left[(1+\beta)^{5/3}(1+S_n)^{5/3}+(1+\beta)^{5/3}(1-S_n)^{5/3} \right. \nonumber \\
\left.+(1-\beta)^{5/3}(1+S_p)^{5/3}+(1-\beta)^{5/3}(1-S_p)^{5/3}\right] \ ,
\end{array}
\label{eq:kinetic}
\end{equation}
being $k_F=(3\pi^2\rho/2)^{1/3}$ the Fermi momentum of nonpolarized isospin 
symmetric matter. 
An idea of the possible terms appearing in the potential energy contribution 
can be extracted from our previous phase space analysis of the single-particle 
potentials. We start from the BHF approximation of the potential energy per 
particle $V/A$ defined through  Eq.\ (\ref{eq:epp}) and making use of  
Eqs.\ (\ref{eq:nu})--(\ref{eq:pd}) for the single particle 
potential $U_{\tau \sigma}(k)$, we can write 
\begin{equation}
\frac{V}{A} \approx \frac{1}{2A} \sum_{\tau,\sigma}
\sum_{\tau^\prime,\sigma^\prime} \sum_{k \leq k_F^{\tau\sigma}} 
g_{\tau\sigma\tau^\prime\sigma^\prime}(k,\rho)\rho_{\tau^\prime \sigma^\prime} \ .
\label{eq:pe1}
\end{equation}
Next we use Eq. (\ref{eq:g1234}) and perform the average of the quantities 
$g_i(k,\rho)$ over the corresponding Fermi sphere with radius 
$k_F^{\tau \sigma}$. 
Finally, we arrive, after some little algebra, to
\begin{equation}
\begin{array}{c}
\displaystyle{\frac{V}{A} \approx \frac{\rho^2}{4}[\bar{g}_1+\bar{g}_2+\bar{g}_3+\bar{g}_4] + 
\frac{\rho^2}{4}[\bar{g}_1+\bar{g}_2-\bar{g}_3-\bar{g}_4]\beta^2} \nonumber \\
\displaystyle{\frac{\rho^2}{8}[\bar{g}_1-\bar{g}_2](1+\beta)^2S_n^2+\frac{\rho^2}{8}[\bar{g}_1-\bar{g}_2](1-\beta)^2S_p^2
+\frac{\rho^2}{4}[\bar{g}_3-\bar{g}_4](1-\beta^2)S_nS_p} \ ,
\end{array}
\label{eq:pe2}
\end{equation} 
where $\bar{g}_i$ are just the averages values of the quantities $g_i$. Following this simple analysis we can finally infer the form of the total energy 
per particle 
\begin{equation}
\begin{array}{c}
\displaystyle{\frac{E}{A}}(\rho,\beta,S_n,S_p)=\displaystyle{\frac{T}{A}}(\rho,\beta,S_n,S_p) 
+V_0(\rho)+V_1(\rho)\beta^2+V_2(\rho)(1+\beta)^2S_n^2 \nonumber \\
+V_2(\rho)(1-\beta)^2S_p^2
+V_3(\rho)(1-\beta^2)S_nS_p \ .
\end{array}
\label{eq:fit}
\end{equation}

This parametrization is consistent with the spin and isospin structure of the nucleon-nucleon interaction, in the sense 
that if we consider a particular configuration, $\beta, S_n, S_p$, of the system, a global flip of the spins does not change the energy, 
whereas a flip only of all neutron or protons spins does. It is also clear that due to charge symmetry, a system with a configuration, $\beta'=-\beta, 
S_n'=S_p, S_p'=S_n$, will have the same energy as the original one (note that the coefficients of the terms $(1+\beta)^2S_n^2$ and $(1-\beta)^2S_p^2$ 
are the same).

The coefficients $V_0(\rho), V_1(\rho), V_2(\rho)$ and $V_3(\rho)$, whose density dependence is shown in Fig.\ \ref{fig:fig5}, have been
determined in the following way 
\begin{equation}
V_0(\rho)=\displaystyle{\frac{V}{A}(\rho,\beta=0,S_n=0,S_p=0)} \ , 
\label{eq:coeff0}
\end{equation}
\begin{equation}
V_1(\rho)=\displaystyle{\frac{V}{A}(\rho,\beta=1,S_n=0,S_p=0)-V_0(\rho)} \ , 
\label{eq:coeff1}
\end{equation}
\begin{equation}
V_2(\rho)=\frac{V}{A}(\rho,\beta=0,S_n=1,S_p=0)-V_0(\rho) \ , 
\label{eq:coeff2}
\end{equation}
\begin{equation}
V_3(\rho)=\frac{V}{A}(\rho,\beta=0,S_n=1,S_p=1)-V_0(\rho)-2V_2(\rho) \ . 
\label{eq:coeff3}
\end{equation}
It is clear, however, that their determination is not unique. Choosing them in this way, we get a parametrization which reproduces with a good quality 
(see Figs.\ \ref{fig:fig6} and \ref{fig:fig7} and the discussion below) the results of the BHF calculation of the total energy per particle for values of 
$\beta, S_n$ and $S_p$ around their values for nonpolarized symmetric matter (i.e., $\beta=0, S_n=0, S_p=0$). 

Note that for nonpolarized matter this 
parametrization reduces to 
\begin{equation}
\displaystyle{\frac{E}{A}}(\rho,\beta,0,0)=\displaystyle{\frac{T}{A}}(\rho,\beta,0,0)
+V_0(\rho)+V_1(\rho)\beta^2 \ ,
\label{eq:fit2}
\end{equation}
which can be identified with the usual parabolic approach of the nuclear matter energy per particle if the kinetic energy 
contribution is expanded up to order $\beta^2$.

In order to test the quality of this parametrization, we show in Fig.\ \ref{fig:fig6} the total energy per particle at $\rho=0.8$ fm$^{-3}$
as a function of the proton spin polarization for different values of the neutron spin polarization and two values of the asymmetry parameter:
$\beta=0$ (right panel) and $\beta=0.25$ (left panel). Circles, squares, diamonds and triangles show the results obtained from the BHF 
calculation, whereas those obtained from the parametrization are reported by solid lines. As can be seen from the figure, the dependence on the spin 
polarizations and the asymmetry parameter predicted by Eq. (\ref{eq:fit}) is well confirmed from the microscopic results. It is interesting to note that 
for a fixed value of $\beta$ and $S_n$ (with $S_n \neq 0$), the minimum of the energy happens for a value of $S_p \neq 0$. However, this is not an 
indication of a phase transition to a ferromagnetic state, because the real ground state of the system, as we have seen, is that of nonpolarized 
matter in all the range of densities and isospin asymmetries considered. For completeness, we compare in Fig.\ \ref{fig:fig7} the results for the total 
energy per particle as a function of density for three arbitrarily asymmetric and spin polarized situations: $\beta=0.25, S_n=0.3, S_p=0.4$, $\beta=0.5, 
S_n=0.5$, $S_p=0.25$, and $\beta=0.8, S_n=0.6$, $S_p=0.2$, obtained from the BHF calculation and from the parametrization. As in the previous figure, 
symbols show results obtained from the BHF calculation, while soid lines refer to those obtained with the parametrization.  The quality of the 
parametrization is quite good, as can be seen in both figures, with deviations from the microscopic calculation of at most $7-8\%$ only for the 
combinations of $\beta, S_n$ and $S_p$ with the largest values. Higher deviations are found, however, at higher values of these parameters, being 
this an indication that the coefficients need to be refitted when considering matter with $|\beta| \sim 1, |S_n| \sim 1$ and $|S_p| \sim 1$.

\subsection{Magnetic susceptibility}
\label{sec:sec3.3}

The ratio det($1/\chi$)/det($1/\chi_F$) can be evaluated in a very simple analytic way from Eq. (\ref{eq:det})
if the parametrization of Eq. (\ref{eq:fit}) is assumed, giving
\begin{equation}
\frac{\mbox{det}(1/\chi)}{\mbox{det}(1/\chi_F)}
=1+\frac{12mV_2(\rho)}{\hbar^2k_F^2}[(1+\beta)^{1/3}+(1-\beta)^{1/3}]
+\frac{36m^2}{\hbar^4k_F^4}[4V_2^2(\rho)-V_3^2(\rho)](1-\beta^2)^{1/3}
 \ ,
\label{eq:det2}
\end{equation}
with $m$ the average mass of the nucleons. This ratio is shown in Fig.\ \ref{fig:fig8} as a function of the density for 
several values of the asymmetry parameter $\beta$ from neutron to symmetric matter. Note that an important increase of the ratio happens 
as soon as a small fraction of protons is introduced in the system. This can be understood by looking at Eq.\ (\ref{eq:det2}) and Fig.\ \ref{fig:fig5}
from which it is clear that the ratio det($1/\chi$)/det($1/\chi_F$) increases as the asymmetry parameter $\beta$ decreases from $1$ to $0$. According to 
the criteria for the appearance of a ferromagnetic instability, such an instability should appear when det($1/\chi$)/det($1/\chi_F$)$=0$. 
Nevertheless, it can be seen from the figure that det($1/\chi$)/det($1/\chi_F$) increases monotonously with density, and a decrease is not to be
expected even at higher densities for any value of the asymmetry parameter. Therefore, it can be inferred from these results that there is no sign at
any density of a possible ferromagnetic phase transition for any asymmetry of nuclear matter. 

As we said  previously, the coefficients of the parametrization should be refitted when large values of the asymmetry parameter and the spin 
polarizations are considered. In particular, for the case of pure neutron matter it is more convenient to redefine coefficient $V_2(\rho)$ in the 
following way
\begin{equation}
V_2(\rho)=\frac{1}{4}\left(\frac{V}{A}(\rho,\beta=1,S_n=1,S_p=0)-\frac{V}{A}(\rho,\beta=1,S_n=0,S_p=0)\right) \ ,
\label{eq:coeff2_bis}
\end{equation}
in order to get a better value of the magnetic susceptibility. We show in the figure by open circles the resulting neutron magnetic susceptibility 
obtained by refitting coefficient $V_2(\rho)$ according to Eq. (\ref{eq:coeff2_bis}). We find a good agreement with the results of ref. \cite{vi02} 
where this coefficient is fitted in this way. For comparison, we show also the results for neutron matter obtained recently by Fantoni {\it et al.} 
\cite{fa01} (filled circles). We note here, that our results confirm the reduction of about a factor $3$ of the magnetic susceptibility of neutron matter 
with respect to its Fermi gas value found by these authors, being the differences ascribed to the method (AFDMC) and the potential  (AU6' two-body 
+ UIX three-body) employed by them. It seems, therefore, that this reduction is largely independent of the two-body modern potential used, and, 
in addition, that the effects of the three-nucleon interaction on the magnetic susceptibility are not large.


\section{Summary and Conclusions}
\label{sec:sec4}

Employing a realistic modern nucleon-nucleon interaction (NSC97e) we have studied properties of spin polarized isospin asymmetric nuclear 
matter within the Brueckner--Hartree--Fock approximation. We have determined the single particle potentials of neutrons and protons with 
spin up and down for several values of the neutron and proton spin polarizations and the asymmetry parameter. We have found that the potentials 
exhibit an almost linear and symmetric variation as a function of these parameters. Deviations from this behaviour occur at higher values
of the asymmetry parameter and spin polarizations. These deviations have to be attributed to the dependence of the nucleon-nucleon 
$G$ matrices on $\beta, S_n$ and $S_p$. 

We have calculated the total energy per particle as a function of the density for totally polarized and nonpolarized neutron, asymmetric and 
symmetric matter. We have found that in the range of densities explored (up to $7\rho_0$) totally polarized matter is always more repulsive 
than nonpolarized matter for any asymmetry. This is due to a combination effect of the kinetic and potential energy contributions.
We have also found in the totally polarized case that asymmetric and symmetric matter is more repulsive when neutron and proton spins are 
antiparallelly orientated than when all the spins are aligned along the same direction.

We have constructed an analytic parametrization of the total energy per particle as a function of the asymmetry parameter and 
spin polarizations. The quality of this parametrization has been tested, finding deviations from the microscopic calculation of at most $7-8\%$. 
Nevertheless, deviations are higher for higher values of the asymmetry parameter and spin polarization, and the coefficients of the parametrization
need to be refitted when large values of the spin and isospin asymmetries are considered. 

Employing this parametrization we have determined the magnetic susceptibility of nuclear matter for several values of the asymmetry from neutron 
to symmetric matter in terms of the ratio det($1/\chi$)/det($1/\chi_F$). We have found that this quantity increases monotonously with density, from 
which it can be inferred that a phase transition to a ferromagnetic state is not to be expected in nuclear matter at any density for any asymmetry. 

Finally, our results confirm the reduction of about a factor $3$ of the magnetic susceptibility of neutron matter with respect to its Fermi gas value
found recently by Fantoni {\it et al.} \cite{fa01}.


\section*{Acknowledgements}

The authors are very grateful to professors A. Fabrocini, A.\ Polls, A.\ Ramos and S. Rosati for useful discussions and comments. One of the authors (I.V.) 
wishes to acknowledge support from the Istituto Nazionale di Fisica Nucleare (Italy).



\begin{table}
\caption{Contribution of the spin singlet and triplet channels to the potential energy per particle of neutron and symmetric matter at 
$\rho=0.17$ fm$^{-3}$. Second and third columns show results for nonpolarized and totally polarized neutron matter, respectively, whereas those for 
nonpolarized and totally polarized symmetric matter are reported on the fourth, fifth and sixth ones. Partial waves have been included up to 
total angular momentum $J=4$.}
\bigskip
\bigskip
\begin{tabular}{c| cc |ccc}
& \multicolumn{2}{c|}{Neutron matter} & \multicolumn{3}{c}{Symmetric matter} \cr 
& $S_n=0$ & $S_n=\pm 1$ \phantom{cac} & $S_n=S_p=0$ \phantom{cac} & $S_n=S_p=\pm 1$ \phantom{cac} & $S_n=\pm 1, S_p=\mp 1$ \phantom{cac} \cr
\hline
$S=0$ \phantom{cac} & $-21.970$ & $-    $ \phantom{cac} & $-14.414$   \phantom{cacaa} & $-      $  \phantom{cacaa} & $-0.616$    \phantom{cacaa} \cr
$S=1$ \phantom{cac} & $-0.247 $ & $9.960$ \phantom{cac} & $-24.650$   \phantom{cacaa} & $-23.173$  \phantom{cacaa} & $-7.647$    \phantom{cacaa} \cr
\hline
Total   \phantom{cac} & $-22.217$ & $9.960$ \phantom{cac} & $-39.064$   \phantom{cacaa} & $-23.173$        \phantom{cacaa} & $-8.263$    \phantom{cacaa} 
\cr
\end{tabular}
\label{tab:tab1}
\end{table}

\begin{table}
\caption{As in Table \ref{tab:tab1} but for $\rho=0.4$ fm$^{-3}$.}
\bigskip
\bigskip
\begin{tabular}{c| cc |ccc}
& \multicolumn{2}{c|}{Neutron matter} & \multicolumn{3}{c}{Symmetric matter} \cr 
& $S_n=0$ & $S_n=\pm 1$ \phantom{cac} & $S_n=S_p=0$ \phantom{cac} & $S_n=S_p=\pm 1$ \phantom{cac} & $S_n=\pm 1, S_p=\mp 1$ \phantom{cac} \cr
\hline
$S=0$ \phantom{cac} & $-37.222$ & $-    $  \phantom{cac} & $-22.619$   \phantom{cacaa} & $-      $  \phantom{cacaa} & $8.291 $      \phantom{cacaa} \cr
$S=1$ \phantom{cac} & $10.090 $ & $48.173$ \phantom{cac} & $-33.102$   \phantom{cacaa} & $-20.014$  \phantom{cacaa} & $3.413$        \phantom{cacaa} \cr
\hline
Total \phantom{cac} & $-27.132$ & $48.173$ \phantom{cac} & $-55.721$   \phantom{cacaa} & $-20.014$        \phantom{cacaa} & $11.704$  \phantom{cacaa} 
\cr
\end{tabular}
\label{tab:tab2}
\end{table}


\begin{figure}[hbtp]
 \setlength{\unitlength}{1mm}
   \caption{Neutron ($n$$\uparrow$,$n$$\downarrow$) and proton ($p$$\uparrow$, $p$$\downarrow$) single-particle potentials as a function of 
the linear momentum $k$ at $\rho=0.17$ fm$^{-3}$ for several values of $\beta$, $S_n$ and $S_p$. Left and middle panels show results for 
symmetric matter, whereas those for asymmetric matter are reported on the right one. Solid lines show as a refence the results for the  
nonpolarized case, while dotted ones refer to the spin polarized cases.}
   \label{fig:fig1}
\end{figure}

\begin{figure}[hbtp]
 \setlength{\unitlength}{1mm}
   \caption{Neutron ($n$$\uparrow$ and $n$$\downarrow$) and proton ($p$$\uparrow$ and $p$$\downarrow$) single-particle
    potentials at $k=0$ and $\rho=0.17$ fm$^{-3}$ as a function of neutron (left panels) and proton (right panels) spin 
    polarizations. Upper (lower) panels show results for symmetric (asymmetric) mattter. $S_n$ ($S_p$) is taken equal to $0$ on the 
    right (left) panels.}
   \label{fig:fig2}
\end{figure}

\begin{figure}[hbtp]
 \setlength{\unitlength}{1mm}
   \caption{Total energy per particle as a function of the density for nonpolarized (solid lines) 
and totally polarized (dotted lines) neutron (left panel), asymmetric (middle panel) and symmetric (right panel) matter.}
   \label{fig:fig3}
\end{figure}

\begin{figure}[hbtp]
 \setlength{\unitlength}{1mm}
   \caption{Kinetic (upper panels) and potential (lower panels) energy contributions to the total energy per particle as a function 
of the density for nonpolarized polarized (solid lines) and  totally polarized (dotted lines) neutron (left panel), asymmetric (middle 
panel) and symmetric (right panel) matter.}
   \label{fig:fig4}
\end{figure}

\begin{figure}[hbtp]
 \setlength{\unitlength}{1mm}
   \caption{Density dependence of the coefficients of the parametrization defined in Eq. (\ref{eq:fit}).}
   \label{fig:fig5}
\end{figure}

\begin{figure}[hbtp]
 \setlength{\unitlength}{1mm}
   \caption{Total energy per particle at $\rho=0.8$ fm$^{-3}$ as a function of the proton spin polarization for different values of the neutron 
spin polarization and two values of the asymmetry parameter: $\beta=0$ (right panel) and $\beta=0.25$ (right panel). Circles, squares, diamonds and 
triangles show the results of the BHF calculation, whereas solid lines refer to the parametrization defined in Eq. (\ref{eq:fit}).}

   \label{fig:fig6}
\end{figure}

\begin{figure}[hbtp]
 \setlength{\unitlength}{1mm}
   \caption{Total energy per particle as a function of the density for three arbitrarily spin polarized 
asymmetric situations: $\beta=0.25, S_n=0.3, S_p=0.4$, $\beta=0.5, S_n=0.5, S_p=0.25$, and $\beta=0.8, S_n=0.6, S_p=0.2$. As in Fig.\ \ref{fig:fig6},
symbols show the BHF results, whereas solid lines correspond the parametrization defined in Eq. (\ref{eq:fit}).}
   \label{fig:fig7}
\end{figure}

\begin{figure}[hbtp]
 \setlength{\unitlength}{1mm}
   \caption{Ratio between the determinats of the matrices $1/\chi$ and $1/\chi_F$ as a function of the density for several values 
of the asymmetry parameter $\beta$ from neutron to symmetric matter.}
   \label{fig:fig8}
\end{figure}

\end{document}